\documentclass[twocolumn,showpacs,preprintnumbers,amsmath,amssymb]{revtex4}

\usepackage{graphicx}
\usepackage{dcolumn}
\usepackage{bm}
\usepackage{multirow}

\begin{document}

\preprint{APS/123-QED}

\title{Room-temperature magnetoresistance switching of Py thin films induced by Fe-nanoparticles grown by STM-assisted CVD}
\author{Jens M\"uller}
 \email{mueller@cpfs.mpg.de}
 \altaffiliation[Present address: ]{Max-Planck-Institute for Chemical Physics of Solids, Dresden, Germany}
\author{Stephan von Moln\'{a}r}
\affiliation{Center for Materials Research and Technology (MARTECH),\\
Florida State University, Tallahassee, FL 32306-4351, USA}
\author{Steffen Wirth}
\affiliation{Max Planck Institute for Chemical Physics of Solids, N\"othnitzer-Str.\ 40, D-01187 Dresden, Germany}

\date{\today}

\begin{abstract}
Arrays of Fe-nanoparticles grown by STM-assited CVD have been placed on top of a narrow stripe of Py. 
The magnetic coupling between the nanoparticles and the underlying Py film results in distinct negative jumps of the Py magnetoresistance. The switching of the magnetization orientation of individual particles is clearly reflected in the Py magnetoresistance as a consequence of AMR and DWMR, with a homogeneous particle magnetization orientation yielding the highest resistances.
\end{abstract}

\pacs{73.50.Jt,75.75.+a}
\maketitle

Ordered magnetic nanostructures  \cite{Cowburn2000,Bader2006} are of continuing interest for application in high-density magnetic storage, magnetic sensing, spintronics, and biology. Along these lines, artificial hybrid systems in which the magnetic nanostructures interact with semiconducting, metallic, superconducting or even magnetic layers (see \cite{Martin2003} and references therein) have recently attracted considerable attention \cite{Liu2002,Pierce2004,Sapozhnikov2007,Villegas2007}. 
Here, we report on hybrid structures consisting of small arrays of magnetic iron (Fe) nanoparticles grown onto an underlying soft magnetic permalloy (Py) thin film. Our approach was to link the magnetization of the individual neighboring particles to one another through the Py layer, such that the underlayer "guides" the particles' flux, i.e.\ enhances magnetic interaction between them. The investigation of such systems is not only driven by the quest for a more detailed picture of the magnetization behavior of the interacting particles themselves, but also to apply these small and local magnetic flux sources to intentionally influence and investigate other materials and thus to observe new effects, e.g., in the transport properties of the magnetic layer. Our hybrid systems show pronounced negative jumps in magnetoresistance (MR) up to room temperature as the magnetization state of the particle array changes: the magnetotransport switches in steps between high-resistance and low-resistance states corresponding to homogeneous and inhomogeneous magnetization configurations of the array, respectively.

The iron particle arrays were grown by a combination of chemical vapor deposition (CVD) and scanning tunneling microscopy (STM) as described in detail elsewhere \cite{Wirth2005}. This method has successfully been used to fabricate particles on various conducting substrates from $5 - 20$\,nm in diameter and $50 - 250$\,nm in height and with interparticle distances down to 80\,nm. 
The investigation of the properties of such elongated cylinders could be of interest for perpendicular magnetic recording due to their shape anisotropy. An advantage of this fabrication technique is the possibility for steering the STM tip so as to 
position the particles exactly with respect to each other and to any feature on the substrate.
\begin{figure}[b]
\includegraphics[width=0.475\textwidth,clip]{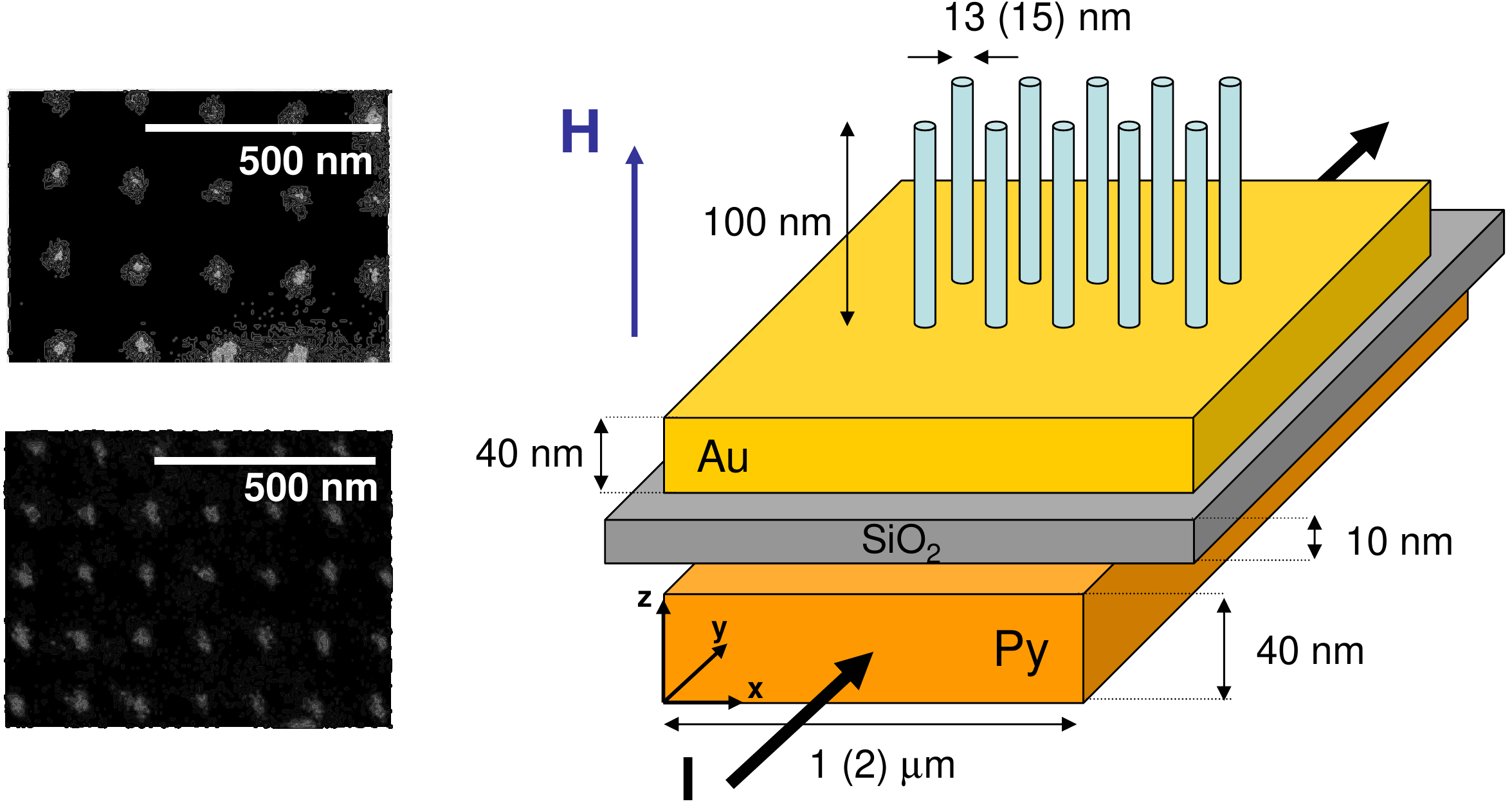}
\caption[]{(Color online) Left: SEM picture of typical arrays of Fe magnetic nanoparticles grown by STM-assisted CVD onto Nb (top) and Au (bottom) substrates. Right: Schematics (not to scale) of the Fe-nanoparticle/Py-thin-film hybrid structures with dimensions for devices A (B). Magnetic field $H$ and electrical current $I$ are applied along the $z$- and $y$-directions, respectively.}\label{figure1}
\end{figure}
Here, the Fe-nanoparticles have been grown on top of two multilayer devices  consisting of a semi-insulating GaAs substrate, a Py film, an insulating SiO$_2$ layer, and an Au top layer, see Fig.\,\ref{figure1} for dimensions. The Py films are lithographically patterned stripes 
with side legs for four-terminal transport measurements. The SiO$_2$ layer (grown by rf-sputtering) isolates spacially the Fe-nanocylinders grown onto the top Au layer from the underlying Py film. As a consequence the interaction between the Fe-nanoparticles and the Py film is solely of magnetostatic nature. These structures allow the study of the transport properties of the magnetic Py film as a function of the magnetization configuration of the Fe-nanoparticle array on top. The metallic Au and Py (Fe$_{0.79}$Ni$_{0.21}$) films  were deposited by thermal evaporation and magnetron sputtering, respectively. The surface roughness of the Py films was about 10\,\AA\ and the average grain size $\sim 25$\,nm. 

\begin{figure}[b]
\includegraphics[width=.425\textwidth,clip]{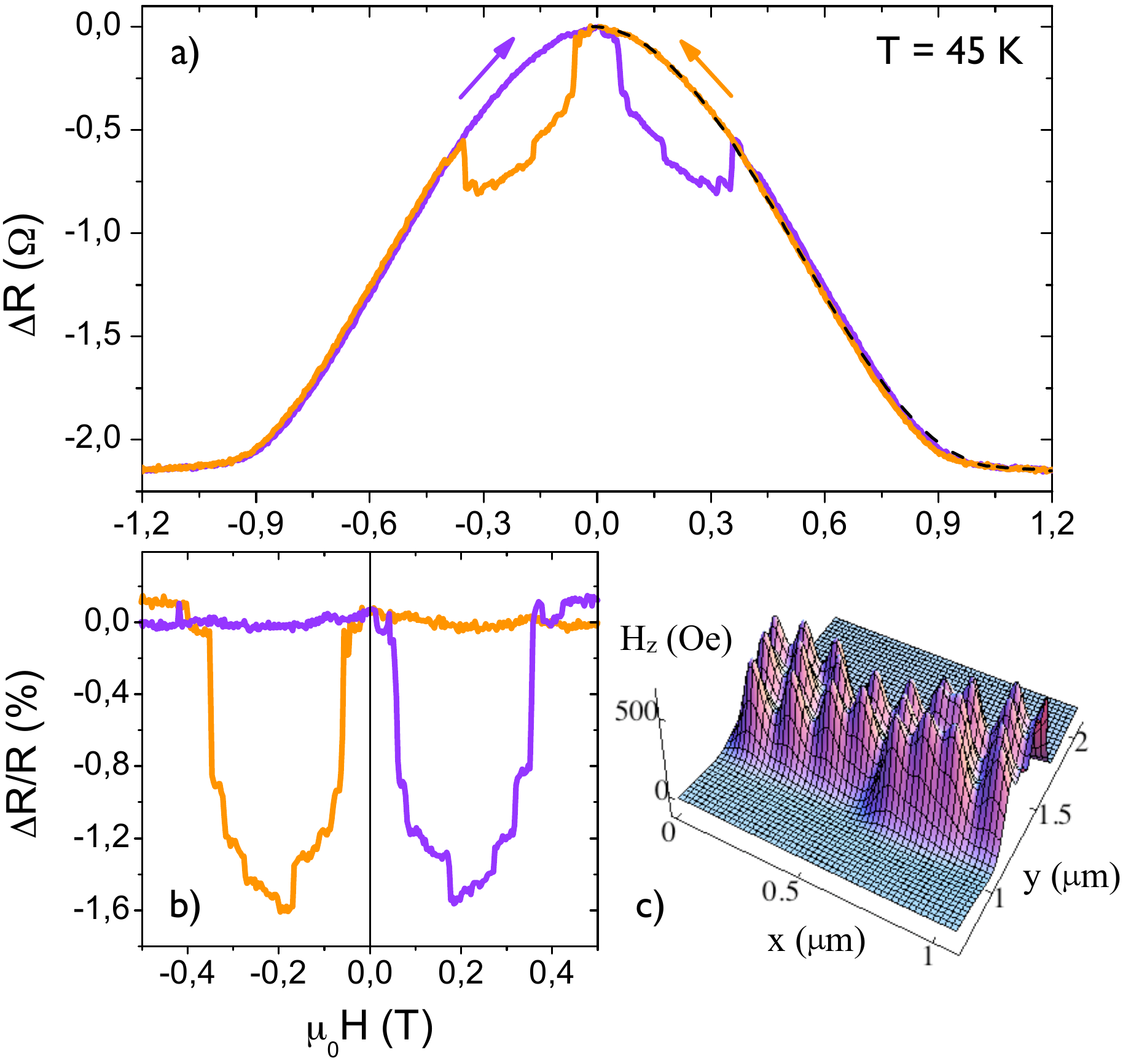}
\caption[]{(Color online) a) Magnetoresistance of device A at $T = 45$\,K. The Fe-nanoparticle array is of rectangular shape ($3 \times 10$, interparticle distance 120\,nm). b) Relative change in MR after a smooth background represented by the dashed line in a), see text, has been subtracted. c) $z$-component of the particle array's stray field $H_z (x,y)$ 75\,nm below the array.}\label{figure2}
\end{figure}
Fig.\,\ref{figure2}\,a) shows the magnetoresistance of device A for magnetic fields applied perpendicular to the Py film, i.e.\ parallel to the Fe-nanoparticles' easy magnetization direction. As expected for this geometry \cite{Rijks1997} the zero-field state is a high resistance state, as the magnetization $M_{\rm Py}$ is in the film plane. In saturation (at about $\pm 1.05$\,T, determined by demagnetization) the magnetization is perpendicular to the applied current $I$ resulting in a low resistance \cite{McGuire1975}. The dashed line nicely fits the MR varying as $\rho(\Theta) = \rho_\perp + \Delta \rho \cos^2 \Theta$, where $\Delta \rho = \rho_\parallel - \rho_\perp$ denotes the difference between the resistivities parallel and perpendicular to $M_{\rm Py}$, and $\Theta$ is the angle between $I$ and $M_{\rm Py}$, as expected for the anisotropic magnetoresistance (AMR) when the magnetization rotates coherently \cite{McGuire1975}. 
The interaction with the Fe-nanoparticle array on top, however, gives rise to a remarkably clear \emph{negative} switching effect with hysteresis. 
Clearly, the hysteresis in MR of the Py film correlates with the one of the $M$-$H$ loop of the particles \cite{Mueller2008,Wirth1999}. This confirms that the former is caused by the particle array's {\em magnetization orientation} rather than $H$.
Shown in Fig.\,\ref{figure2}\,b) is the relative resistance change after the background has been subtracted. Clear steps indicate switching of individual or groups of few particles within the array.\\ 
The dipolar field created by one particle acting on the neighboring one is opposite to its magnetization direction. The enhanced interaction mediated by the magnetic film leads to a stabilization of certain configurations of the magnetic moments of neighboring particles (as observed by MFM at room temperature) and a broadened switching distribution to minimize stray field energy \cite{Wirth2000,Christoph2001}. 
We find that the zero-field resistance of the Py film takes on various different values depending on the sweep history, and thus strongly depends on the actual magnetization configuration of the Fe-nanoparticle array. For magnetic fields applied parallel to the film (in-plane saturation field $\sim 0.05$\,T) no switching effect in the permalloy MR has been observed. Obviously, for this field configuration the Fe particle magnetization merely rotates coherently \cite{Wirth1999}. 
\begin{figure}[b]
\includegraphics[width=0.425\textwidth,clip]{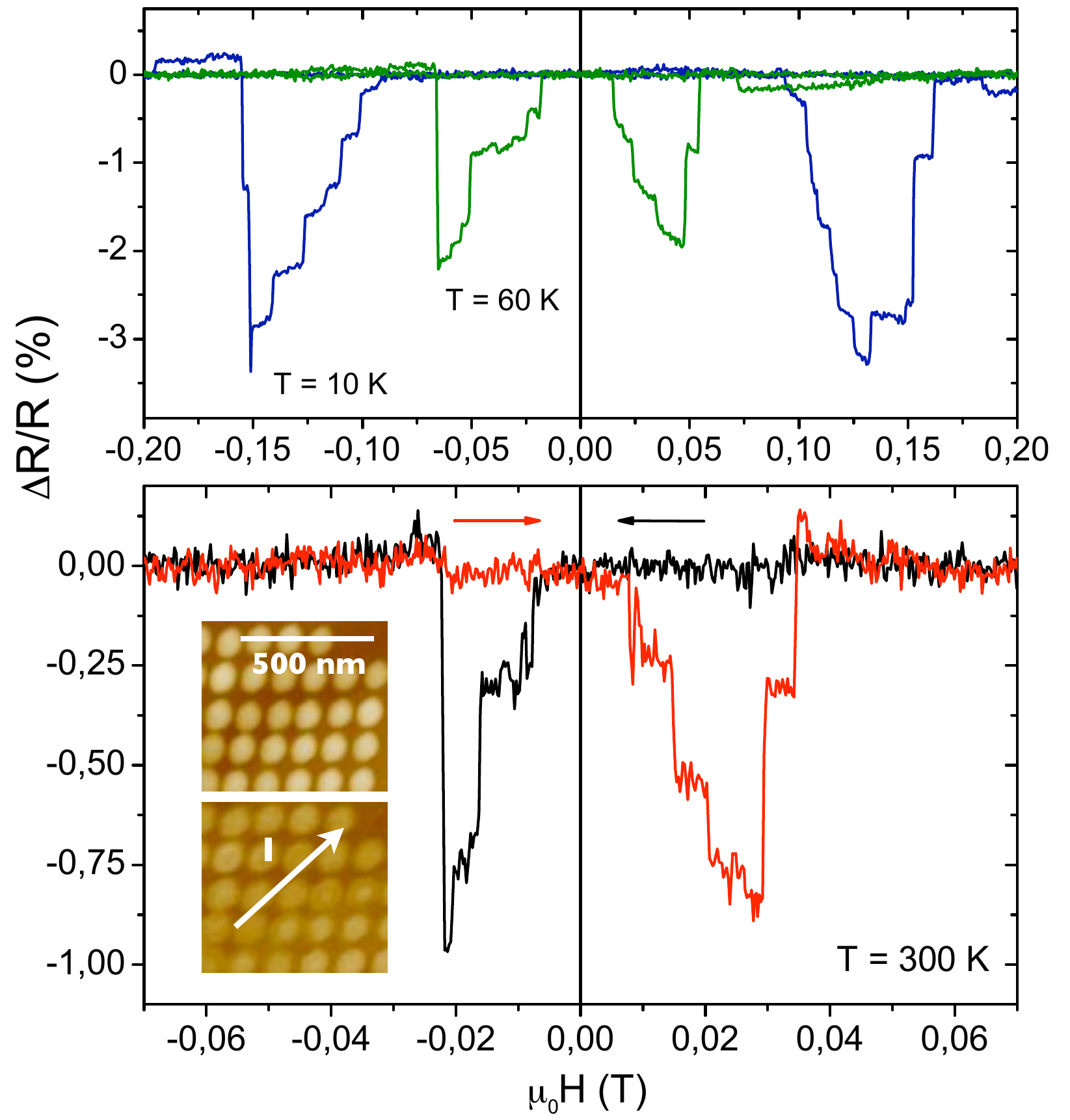}
\caption[]{(Color online) MR curves of device B ($5 \times 11$ array, interparticle distance 120\,nm) taken at different temperatures (background subtracted). Inset: room-temperature AFM (top) and MFM (bottom, at $H=0$) of the Fe-nanoparticle array (edges of the Py stripe not shown).}\label{figure3}
\end{figure}\\
In Fig.\,\ref{figure3}, results on a second device B are shown at different temperatures. Room-temperature AFM and MFM measurements reveal the structure of the array and confirm a stable magnetization orientation at 300\,K \cite{Wirth2001}. The hysteresis in MR at different temperatures again reveals step-like negative switching events: several distinct jumps can clearly be resolved. Remarkably, even at room temperature distinct steps can be resolved indicating magnetization reversal of groups of particles within the array.
MFM data taken after a field pulse of about 20\,mT (not shown), i.e.\ in the low MR state, confirm an inhomogeneous configuration with lines of particles with parallel magnetization orientation and adjacent lines having opposite magnetization orientations similar to prior observations \cite{Wirth2000}. Obviously, the negative switching in the Py MR occurs then when a certain inhomogeneous magnetic moment configuration of the Fe-nanoparticle array is stabilized.\\ 
Fig.\,\ref{figure4} displays the temperature dependence of the observed switching fields $H_{\rm sw}$ (left) and the relative resistance change $\Delta R/R$ (right) for devices A and B. 
%
For device A (see inset) the switching distribution is substantially broader than for device B indicating particularly strong interparticle interactions as expected for the smaller interparticle distance of the former. 
The temperature dependence of $H_{\rm sw} (T)$ which differs from the linear behavior of noninteracting particles \cite{Wirth1999} as well as the strong decrease of $\Delta R/R$ with increasing temperature probably reflects the complex hybrid nature of the present devices.
\begin{figure}[t]
\includegraphics[width=0.4\textwidth,clip]{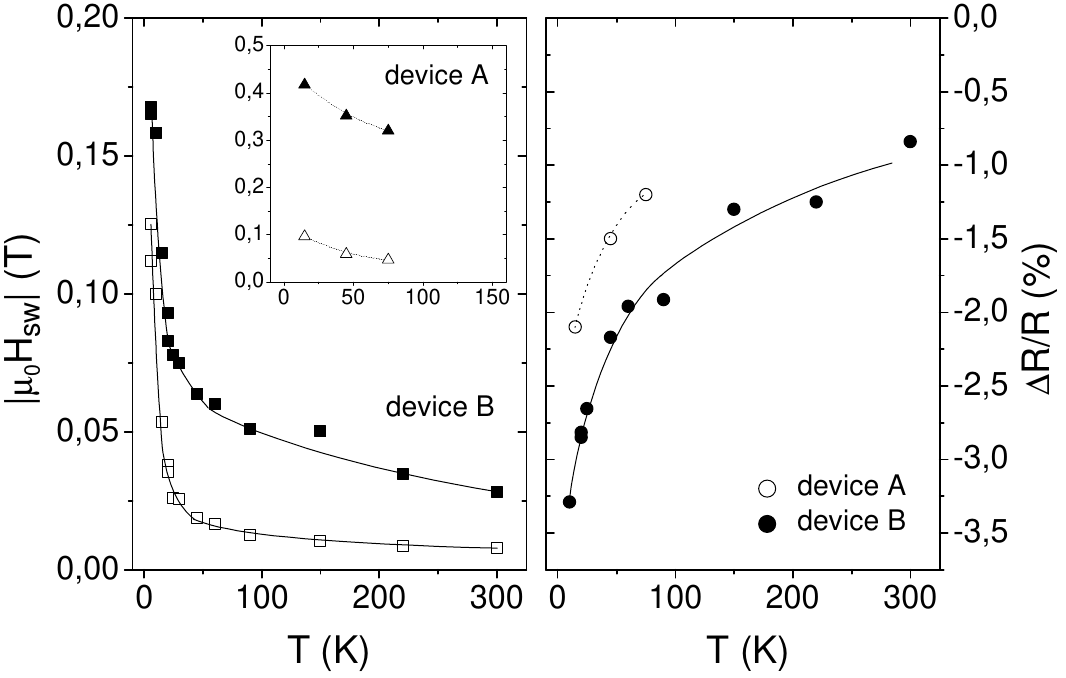}
\caption[]{Temperature dependence of the maximum and minimum switching fields in MR (left) and the magnitude of the relative resistance change (right) for devices A and B. Lines are guides to the eye.}\label{figure4}
\end{figure}

We have shown that the transport properties of a narrow Py film can be controlled by manipulating the magnetic configuration of an array of Fe-nanoparticles in close proximity. 
We find that an inhomogeneous, i.e.\ {\em more disordered}, magnetization configuration of the Fe-NP array corresponds to a {\em lower resistance} state of the Py film. If a mere stray field effect was taking place, a shift of the MR curves along the field axis was expected, which is not observed. We, therefore, consider AMR: in such a scenario a stray field $H_z (x,y)$ of the Fe-nanoparticle array acts such that it rotates $M_{\rm Py}$ out of the Py-film plane (and current direction) leading to a decrease in resistance. If all particles in the array have parallel (either all up or all down) magnetization, $H_z (x,y)$ does not change sign at the Py film. Average values are about 350\,Oe with oscillations of $\pm 100$\,Oe along the $x$- and $y$-coordinates on the length scale determined by the interparticle distance of about 100\,nm, see Fig.\,\ref{figure2}\,c). Thus, one would only expect a smooth modulation of $M_{\rm Py}$ for a saturated array. If lines or columns of particles reverse their magnetization, however, $H_z (x,y)$ periodically {\em changes sign} on a wider length scale ($\sim 300$\,nm for about half the particles switched) between values of $\pm 400$\,Oe. This modulates the Py magnetization much more strongly and thus may {\em locally} cause $M_{\rm Py}$ to be rotated out of the $xy$-plane leading to a lower resistance. A simple calculation shows that, averaged over the area of the array, $M_{\rm Py}$ would have to be rotated only by about 25$^\circ$ out of the plane to explain the effect displayed in Fig\,\ref{figure2}.
Besides this AMR effect involving a mere modulation in amplitude of $M_{\rm Py}$, it might also be conceivable that the Py film's actual domain structure is modified on length scales comparable to the interparticle distances. 
%
Numerical simulations reveal that, due to the local character of the particle stray fields, the domain patterns are strongly influenced by the Fe-nanoparticles. Extended linear domain walls (DW) are formed in the layers of the particle/film hybrid system (see Fig.\,2 in \cite{Christoph2001}). This numerical result 
suggests the existence of walls (not necessarily of $180^\circ$) inside the permalloy between particle lines with opposite magnetization orientation. The domain walls might add to the observed resistance changes and the DWMR in the present devices would be of negative sign.
Clearly, detailed micromagnetic simulations are desirable to distinguish between the contributions of AMR and DWMR. Nonetheless, our results show that engineering the transport properties of a magnetic thin film by decorating it with individual nano-scale magnetic particle appears to be an intriguing perspective for possible applications, in particular since the effect persists up to room temperature. 



\end{document}